**Shear piezoelectricity in poly(vinylidenefluoride-*co*-trifluoroethylene): full piezotensor coefficients by molecular modeling, biaxial transverse response, and use in suspended energy-harvesting nanostructures**

*Luana Persano*, *Alessandra Catellani*, *Canan Dagdeviren*, *Yinji Ma*, *Xiaogang Guo*, *Yonggang Huang*, *Arrigo Calzolari*, and *Dario Pisignano*

Dr. L. Persano and Prof. D. Pisignano
CNR-NANO, Istituto Nanoscienze, Euromediterranean Center for Nanomaterial Modelling and Technology (ECMT), via Arnesano I-73100, Lecce (Italy)
E-mail: luana.persano@nano.cnr.it
Dr. A. Catellani and Dr. A. Calzolari
CNR-NANO, Istituto Nanoscienze, Centro S3, via Campi 213, I-41125 Modena (Italy)
Dr. C. Dagdeviren
Department of Materials Science and Engineering, Frederick Seitz Materials Research Laboratory, and Beckman Institute for Advanced Science, University of Illinois at Urbana-Champaign, Urbana, IL 61801 (USA)
The David H. Koch Institute for Integrative Cancer Research, Massachusetts Institute of Technology, Cambridge, MA 02139 (USA)
Harvard Society of Fellows, Harvard University, Cambridge, MA 02138 (USA)
Dr. Y. Ma, Dr. X. Guo and Prof. Y. Huang
Department of Civil and Environmental Engineering and Department of Mechanical Engineering, Northwestern University, Evanston, IL 60208 (USA)
Dr. Y. Ma
Department of Engineering Mechanics, Center for Mechanics and Materials, Tsinghua University, Beijing 100084 (China)
Dr. X. Guo
College of Aerospace and Civil Engineering, Harbin Engineering University, Harbin 150001 (China)
Prof. D. Pisignano
Dipartimento di Matematica e Fisica "Ennio De Giorgi", Università del Salento, via Arnesano I-73100 Lecce (Italy)

Keywords: Polymer nanofibers, Electrospinning, Electromechanical coupling, Piezoelectric properties

Piezoelectric materials and associated nanostructures accumulate electrical charges on their surfaces in response to an applied mechanical stress, through the change in their spontaneous electric polarization.[1,2] Exploiting deformations induced by motion, mechanical vibrations, and environmental noise,[3,4] these systems are extremely attractive for energy harvesting in information and communications technologies and personalized electronics.[5-7] Solid-state materials such as crystals[8] and ceramics[9] have been integrated in complex networks for the internet of things[10] as actuators, sensors, and transducers,[11] and as switches in memory devices. Recently, the emerging of magnetoelectric data storage,[12] self-power sources for smart wearables[13] or implantable biomedical devices[14,15] fostered to conjugate mechanical energy harvesting with easy shaped, biocompatible flexible materials.[16,17] In particular, the request of bendable and stretchable systems can be fulfilled with elongated nanostructures (e.g. nanowires and nanotubes).

In this framework, organics show an unequalled processing flexibility, lightweight, large-area and low-cost manufacturing methods, biocompatibility, and low acoustic and mechanical impedance, which make them ideal for underwater and medical applications.[14,15,18] For instance, copolymers of vinylidenefluoride (VDF, $[CH_2-CF_2]_n$) with trifluoroethylene (TrFE), are stable and achieve a high degree of crystallinity (>90%).[19] In addition, they do not need to be poled because they directly crystallize from melt or solution into the ferroelectric ($\beta$-) phase. Piezoelectricity in these materials is related to the electronegativity difference in hydrogen and fluorine atoms, which determines an effective dipole moment in the direction normal to the carbon backbone. Consequently, these films or nanostructures are often utilized with top/bottom contacts.[20-23]

Instead, the special piezoelectric properties of polymers such as the poly(vinylidenefluoride-*co*-trifluoroethylene) [P(VDF-TrFE)] might lead to the development of much more versatile nanogenerator architectures. Differently from crystalline inorganic solids, for which normal piezoelectricity is generally exploited and conveniently achieved by strain along the spontaneous polarization (**P**, **Fig. 1**a), in flexible polymeric systems the stress applied along one axis causes remarkable deformations also along perpendicular directions.[24] This effect, along with the reduced alignment of the polymeric chains and the presence of glassy grains, strips out the concept of uniaxial piezoelectricity and more complex '*transverse*' contributions are to be taken into account, in polarization and in the undergone distortions. Most studies assume only uniaxial models for polymers and organic-inorganic nanocomposites,[12,25] which may result in a limiting description of the response because of the assignment of a reduced number of piezoelectric parameters, as pertaining to systems much more symmetric than the actual ones. In this work, we demonstrate the biaxial shear activity in P(VDF-TrFE), which show the presence of two net components of the electronic polarization ($P_x$, $P_y$ according to the geometry displayed in Fig. 1b) in the plane perpendicular to the chains of macromolecules. Findings here reported involve various aspects:

(i) the full $e_{ij}$ piezoelectric tensors and the Born effective charges due to displacement-induced polarization changes are obtained by first-principles molecular simulations, which highlight that *multiple shear components* ($e_{34}$ *and* $e_{35}$) are non-zero in P(VDF-TrFE). Polyvinylidenefluoride (PVDF) is a 1D polar system, with poling direction along the *y* axis, whereas the inclusion of TrFE units strongly increases F-F repulsion and leads to the establishment of a $P_x$ component, i.e. to a biaxial polarization character (Fig. 1c,d);

(ii) The microscopic shear is exploited in single suspended P(VDF-TrFE) nanobeams, deformed by localized bending to generate 20-40 μV peak voltages and exhibiting high robustness and excellent adhesion at contacts;

(iii) The resulting electromechanical behavior is also studied by means of an analytic model through the *ab initio* calculated piezoelectric coefficients, showing that the shear stress makes substantial contributions to the output voltage.

In the following, we report on these findings in the same order introduced above.

Our nanowires are realized by electrospinning (see the Experimental Section for details). This process is based on the elongation of a polymer flow under an intense electric field, leading to flexible filaments with ultra-high length/diameter ratio and generally circular cross-section. These filaments are robust, and they can be bent repeatedly. Their polarization is mainly along a transverse direction in the plane (*x-y* in Fig. 1) perpendicular to the C backbone and to the nanowire longitudinal axis. Hence, the realization of piezoelectric generators and sensors, in which electrical signals are measured along the nanowire length, is related to shear piezoelectricity.

*Ab initio calculations.* The considerations above highlight the importance of a full, microscopic understanding of the piezoelectric tensor, including shear components. To this aim, we perform *ab initio* pseudopotential Density Functional Theory (DFT) calculations of the piezoelectric coefficients for different copolymer configurations. Since electrospun nanowires contain thousands of polymer chains, we safely neglect surface effects and simulate filaments as bulk-like infinite systems. Firstly, the quality of this approximation is estimated by comparing experimental and calculated infrared transmission spectra. **Fig. 2** displays these spectra for fibers (black curves) and the theoretical results (red curves) for both VDF and VDF-TrFE compounds. All the fundamental features of the fibers are properly reproduced by the theoretical results, which supports the validity of the developed model. The discrepancies between the experimental and the calculated curves (such as the different spectral broadening) can be ascribed to the effect of local disorder or thermal fluctuations. For instance, considering intrachain disorder, as that due to the possible relative rotation of the $CF_2$ units around the main polymer axis, may cause broader and more structured spectra (see Supporting Information). Together with other inter- and intra-chain distortions, this finely tailors the vibrational frequencies and linewidths, thus well explaining the observed differences of experimental and calculated spectra.

The picture of the formula units of PVDF and P(VDF-TrFE) in Fig. 1c and 1d, showing the optimized (minimum energy) structures of the two molecules, allows for a direct comparison of the dipole moments and resulting piezoelectric coefficients, as extracted by *ab initio* simulations. Symmetry arguments impose PVDF to have non-zero normal piezoelectric coefficients only along the polar (*y*) axis ($e_{21}$, $e_{22}$, $e_{23}$) and correspondingly shear coefficients for distortions in planes that do not contain this axis ($e_{34}$, $e_{16}$). We find that full atomic relaxation in PVDF does not break the symmetry of the system, giving rise to a uniaxial polarization (|P|= 0.20 C/m$^2$), in agreement with previous theoretical reports.[18] The inclusion of TrFE units along the chain (Fig. 1d), increasing F-F repulsion, imparts in-plane rotation of the monomers around the polymer axis. TrFE units have a net dipole moment, however the misalignment of the monomers along the chain leads to a global reduction of the absolute value of the polarization (|P|= 0.16 C/m$^2$) and, more interestingly, to a non-zero component of the polarization along the *x* direction, i.e. a biaxial polarization character.

The piezoelectric response is obtained via full relaxation of the internal coordinates of the system upon applied normal and shear stress to the supercell parameters. The calculated, $e_{ii}$ normal components of the proper piezoelectric tensor of PVDF and P(VDF-TrFE) for *i*=1-3 are shown in Table 1, together with the diagonal elements $\tilde{k}_{ii}$ of the dielectric tensor, whereas the ($e_{3i}$) shear components are summarized in Table 2. These findings have immediate implications for nanostructures and devices. For perfectly crystalline PVDF nanowires lying along the *z* direction, only the shear coefficient $e_{34}$, corresponding to a solicitation along *y*, would contribute to the piezo-voltage collected by contacts at the opposite ends of the nanostructure. Instead, for a VDF-TrFE copolymer nanowire, one could in principle observe piezoelectric response upon solicitation along any axis in the *x*-*y* plane. Furthermore, with negligible normal coefficients, i.e. with zeroing $e_{33}$, one would expect an insignificant contribution of stretching distortions along the nanowire axis to the overall piezoelectric response.

Since the establishment of a biaxial polarization is induced by the presence of TrFE units, it is interesting to understand how it depends on the distribution of these units along chains. We compare the structure of VDF-TrFE shown in Fig. 1d, where the TrFE units are non-consecutive along the chain (configuration hereafter denoted as 'diluted'), with another VDF-TrFE compound having a similar amount of TrFE units, but consecutively aligned along the chain ('clustered' model). From the calculation of the formation energy of fully relaxed systems, the diluted crystal is found to be more stable by 20 meV per monomer (~0.5 kcal/mol). In addition, distortions of the bonds as well as F-F interactions increase with consecutive TrFE units, and the total spontaneous polarization is significantly reduced (|P|= 0.02 C/m$^2$). Finally, our calculations allow the Born effective charges, $Z^*$, to be obtained, quantifying the polarization changes induced by atomic displacements. Born charges provide a local contribution to changed polarization and implicitly to piezoelectric response, due the atomic relaxation of single atoms in response to the lattice deformation (i.e. applied strain). We find that the inclusion of TrFE units weakly affects the absolute values of $Z^*$. For instance, the values of the Born charges for the $C_H$, the $C_F$ and the F atom are -0.28$e$, +1.49$e$ and -0.82$e$, and -0.30$e$, +1.43$e$ and -0.82$e$ for VDF and VDF:TrFE, respectively, where $e$ is the elemental charge of an electron. Further details on performed calculations and full tensor matrices are provided in the Supporting Information file. Overall, simulations highlight that, due to shear piezoelectricity, bendable nanogenerators and sensors can take advantage of geometries where deformation and response do not necessarily involve the same axis.

*Single nanobeam piezoelectric generator.* In order to exploit this concept, we focus on a device architecture with electrodes collecting a piezo-voltage along a direction perpendicular to the applied stress, performing experiments on individual suspended wires of P(VDF-TrFE) with average diameter of 400-600 nm. Shadow masks are used to collect the wires, which can be suspended over lengths ranging from millimeters to many centimeters on both stiff and flexible substrates carrying aluminum electrodes. The procedure to realize the devices is summarized in **Fig. 3a** and involves the preparation of Al masks to be placed on a rotating collector in an electrospinning set-up, the collection of single spun nanofibers across the gap realized in these masks, and the definition of electrodes by Ag paste at the two edges of suspended fibers. Following the separation of the two metallic pads, the system (Aluminium+fiber+Ag contacts) is then fixed on top of glass slides by using a bi-tape. In previous works,[26,27] piezoelectric nanofibers or nanowires have been often used with an intimate integration with flexible substrates, namely applying strain to those substrates which then transmit a deformation to the deposited nanostructures. Due to the needed mechanical stiffness, reports with suspended structures are largely limited to inorganic materials such as ZnO and lead zirconate titanate,[28,29] and to wires with typical length of tens to few hundreds of micrometers.[28] Here, we study a different approach, in which the strain is applied directly to a single, mm-long suspended polymer nanostructure through localized bending induced by a triboindenter tip. This implies mechanical robustness of the developed material, and optimal adhesion at contacts.

The representative scanning electron microscopy (SEM) images of the suspended wires are displayed in Fig. 3b. The nanowires exhibit a smooth surface and uniform diameter along their entire length, and exceptionally high length/diameter ratio (up to $5\times10^5$) in comparison with polymer nanostructures realized by other methods.[26,30-35] These electrospun materials have been found to exhibit superior piezoelectric performance,[26,36] although their operation is not fully explained by traditional models. To evaluate the piezoelectric response under specific compressive loads, flexible and thin Cu wires with a layer of silver epoxy are used to define contacts to the ends of 9 mm long polymer nanobeams. A lock-in amplifier allows the open-circuit voltage to be collected while the beams undergo well-defined levels of displacement, in sub-microscale three-point bending experiments. Fig. 3c shows the generated piezo-voltages during dynamic loading-unloading cycles. The periodic alternation of negative and positive peaks corresponds to the application and release of the stress. The peak output voltage generated by an individual suspended fiber increases from 20 µV to 40 µV upon increasing the applied deformation (up to 109 nm along the *y*-direction). This evidences voltage generation from single piezoelectric polymer nanowires under a flexural deformation directly applied to the filament.

*Device model exploiting the ab initio calculated piezo-coefficients.* The observed behavior, particularly the transverse piezo-electric voltage generated by shear stresses, can be well explained by an analytic electromechanical model of the suspended nanostructure. **Fig. 4**a and b present a schematic illustration of piezoelectric nanobeam with both ends clamped and subjected to a concentrated eccentric loading. The beam has length *l*, piezoelectric and dielectric constants $e_{ij}$ and $k_{ij}$ [$k_{ij} = \tilde{k}_{ij} \times \varepsilon_0$, where $\varepsilon_0$ indicates the vacuum permittivity)] with *a* indicating the position of the applied force. We embed the $e_{34}$ coefficient obtained by the *ab initio* calculations above in the electromechanical model, thus providing a coherent description of experimental findings. The resulting maximum transverse piezo-electric voltage, $V_\tau$, generated by shear stress (i.e., by taking $e_{33}=0$) is obtained as (see Supporting Information for details):

$$V_\tau = \frac{3e_{34}r^2l(2a-l)(1+\nu)}{a^2(l-a)^2 k_{33}} d \tag{1}$$

where *r* is the radius of the piezoelectric beam and *d* is the displacement at the loading point. Fig. 4c shows the transverse piezo-electric voltage response at different positions of bending point obtained for *l*=9 mm, $\nu$=0.35[37], $k_{33}$=2.1×10$^{-11}$ F/m, $e_{34}$ = 0.12 C/m$^2$, *r*=300 nm, and Published in Advanced Materials. Doi: 10.1002/adma.201506381 (2016).$d$=109 nm. Shear components explicitly appear in this formulation. Results show that the transverse piezo-electric voltage depends strongly on the position of the bending point ($a$) making this system interesting for detecting eccentric loads and for nanoscale position sensing. Smaller $a$ corresponds to higher transverse piezo-electric voltage. Fig. 4d compares the voltage response obtained by experiments and theory for $d$=38 nm, $d$=48 nm and $d$=109 nm. The experimental results (0.02 mV, 0.025 mV and 0.038 mV for $d$=38 nm, 48 nm and 109 nm, respectively) are in the same range of theoretical values for $a$ between 1.5 mm and 2.5 mm.

In summary, by analyzing the shear behavior, the response in suspended geometries can be fully rationalized upon taking into account transverse contributions. Given the wide use and importance of PVDF and P(VDF-TrFE) materials for the realization of nanogenerators, pressure sensors, accelerometers, and similar devices, making the full $e_{ij}$ piezoelectric tensors available by first-principle calculations might be very useful. Indeed, multidirectional components of electronic polarization give rise to piezoelectric contributions for a large number of spatial distortions, providing polymer nanostructures with an enhanced overall piezoelectric response. The resulting architectures are extremely versatile, and they can find application in devices for nanoscale localization nearby large-scale surfaces and in systems for harvesting vibrational noise, which might activate the piezo-electric response of nanobeams without involving large-scale deformation of the underlying supporting substrates. Coupling flexural deformation and transverse piezoelectric response under compression forces establishes new rules for designing next devices based on polymer piezoelectric nanomaterials.

*Experimental Section*

*Nanobeam production and characterization.* P(VDF-TrFE) solutions were prepared as reported in Ref. 24, and placed into a 1.0 mL syringe tipped with a 27-gauge stainless steel needle. A bias of 25 kV was applied to the needle by a high voltage supply (EL60R0.6-22, Glassman High Voltage), and the solution was injected into the needle at 1 mL h$^{-1}$ with a syringe pump (33 Dual Syringe Pump, Harvard Apparatus). A grounded cylindrical collector (diameter 8 cm), was placed at a distance of 6 cm from the needle. Properly designed shadow masks were positioned on the surface of the disk to deposit single nanowires. Nanowires with a measured VDF:TrFE ratio of about 70:30[24] were inspected by SEM with a Nova NanoSEM 450 system (FEI), using an acceleration voltage around 5 kV and an aperture size of 30 μm. Fourier transform infrared (FTIR) spectroscopy was performed with a spectrophotometer (Spectrum 100, Perkin-Elmer Inc.) using a 4 mm wide beam incident orthogonally to the plane of the samples. FTIR measurements were performed in air at ambient conditions (temperature=21°C, humidity= 35%).

A triboindenter TI 950 (Hysistron) equipped with either a cono-spherical diamond tip (100 μm diameter) or a flat-ended cylinder sapphire tip (1 mm diameter) was used to perform load/unload cycles on suspended fibers. A maximum load of 0.5 mN was applied at a rate of 0.1 mN/s with holding time of about 2 s. The Young modulus of the fiber, $E$, is related to the obtained reduced modulus, $E_r$, and to the indenter properties (modulus, $E_{IN}$, and Poisson's ratio, $v_{IN}$) by contact mechanics[38] as $E_r^{-1}=(1-v^2)E^{-1}+(1-v_{IN}^2)\ E_{IN}^{-1}$, where $v$ is fiber the Poisson's ratio. The modulus obtained for the electrospun wire was 2.3 GPa, in agreement with measurements on 300 nm thick-films of the same material.[39] A custom data-recording system consisting of a lock-in amplifier (SR830, Standard Research Systems), a multiplexer (FixYourBoard.com, U802), and a laptop was used to capture open-circuit voltage data from the piezo-wires. An input reference signal of 1 kHz was set. The measurements were carried out at room temperature. Wires were suspended on glass substrates, with air gaps of tens of microns given by the thickness of the aluminum foil supporting their edges. A scheme of the used set-up is reported in Fig. S1 in the Supporting Information.

*First-principles theory.* Structural and electronic properties of VDF and VDF-TrFE systems were obtained by using total-energy and forces minimization, based on PBE-GGA implementation of DFT, as coded in Quantum-Espresso.[40] Ionic potentials were described by *ab initio* ultrasoft pseudopotentials and single-particle wave functions were expanded in a plane-waves basis with an energy cutoff of 30 Ry. Infinite polymeric crystals were simulated using periodically repeated supercells. A (6×6×4) *k*-point grid was used for summations over the 3D Brillouin zone. The optimized geometries of the different systems were obtained relaxing all degrees of freedom in supercells that are multiple of formula units, further corrected via relaxation of cell volume. Supercells included six monomers arranged in two parallel chains, aligned along the *z* direction, as in the predicted crystalline *β*-phase and in agreement with previous theoretical calculations.[41, 42]

A Van der Waals correction (Grimme formulation)[43] to dispersive forces was included to improve the description of distortion intra- and inter-chain interactions. Each structure was fully relaxed, until forces on all atoms become lower than 1 meV/Å. Spontaneous polarization was evaluated within the Berry Phase approach.[44] Piezoelectric coefficients were obtained from finite difference derivation upon applied normal and shear cell deformation up to ±0.5%, with respect to the equilibrium parameters. In particular, the (improper) piezoelectric coefficients ($e_{ilk}=\partial P_i/\partial \varepsilon_{lk}$) were achieved as variations of the spontaneous polarization components ($P_i$) as a function of the strain deformation ($\varepsilon_{lk}$). The so-called proper piezoelectric coefficients are linked to the improper ones by a simple relation which takes into account the branch phase dependence of the polarization vector (see Supporting Information for analytical definitions).[45] Using the Voigt notation for strain indices, $lk \rightarrow j$ [i.e. $11 \rightarrow 1$, $22 \rightarrow 2$, $33 \rightarrow 3$, $23(32) \rightarrow 4$, $13(31) \rightarrow 5$, $12(21) \rightarrow 6$], the complete proper piezoelectric response is described by an order-two tensor $e_{ij} = \partial P_i / \partial \varepsilon_\varphi$ with $i$=1, 2, 3 and $j$=1, …, 6, where $j$=1-3 are related to normal strain and $j$=4-6 to shear strain (Fig. 1c,d). The dielectric tensor, Born charges ($Z^* = -\Omega \partial \mathbf{P}/\partial \mathbf{r}$, where $\Omega$ is the unit cell volume and **r** indicates the coordinate of an atomic displacement), and vibrational properties (i.e. phonon eigenmodes and phonon frequency) were also determined in the regime of linear response, by using a joint finite-differences/finite-fields approaches for solid-state systems.[46] Once known phonon modes and $Z^*$, also the infrared cross-sections could be obtained in the linear response.[47]

*Analytic electromechanical model.* A mechanics model of a beam with both ends clamped was established for the suspended piezoelectric wire. The model gives both the shear and normal strain distributions in the piezoelectric fiber, which together with the electromechanical analysis, provided the voltage response analytically (Supporting Information). The uniquely developed features include taking into account eccentric loads applied on the beam at given positions with respect to the fiber length, as well as the shear piezo-response.

**Acknowledgements**
L.P. and C.D. thank Prof. J. A. Rogers for helpful discussion and continued support. The research leading to these results has received funding from the European Research Council under the European Union's Seventh Framework Programme (FP/2007-2013)/ERC Grant Agreements n. 306357 (ERC Starting Grant "NANO-JETS"). We also acknowledge the EU Marie-Curie Initial Training Network Grant Agreement N. 265073 (ITN-Nanowiring) and the CNR 2015 Adjustment Program at S3/ECMT. Y.M. acknowledges the support from the National Natural Science Foundation of China (Grant No.11402135). Y.H. acknowledges the support from NSF (CMMI-1300846 and CMMI-1400169) and the NIH (Grant No. R01EB019337).

Published in Advanced Materials. Doi: [10.1002/adma.201506381](10.1002/adma.201506381) (2016).[1] J. F. Nye, in *Physical properties of crystals* (Oxford Science Publications 1955).

[2] S. R. Anton, H. A. Sodano, *Smart Mater. Struct.* **2007**, *16*, R1.

[3] C. Dong-Hoon, H. Chang-Hoon, K. Hyun-Don, Y. Jun-Bo, *Smart Mater. Struct.* **2011**, *20*, 125012.

[4] K. K. Korir, G. Cicero, A. Catellani, *Nanotechnology* **2013**, *24*, 475401.

[5] C. Dagdeviren, B. D. Yang, Y. Su, P. L. Tran, P. Joe, E. Anderson, J. Xiab, V. Doraiswamy, B. Dehdashti, X. Feng, B. Lu, R. Poston, Z. Khalpey, R. Ghaffari, Y. Huang, M. J. Slepian, J. A. Rogers, *Proc. Natl. Acad. Sci. USA* **2014**, *111*, 1927.

[6] C. Dagdeviren, Y. Shi, P. Joe, R. Ghaffari, G. Balooch, K. Usgaonkar, O. Gur, P. L. Tran, J. G. Crosby, M. Meyer, Y. Su, R. C. Webb, A. S. Tedesco, M. J. Slepian, Y. Huang, J. A. Rogers, *Nat. Mater.* **2015**, *14*, 728.

[7] C. Dagdeviren, Y. Su, P. Joe, R. Yona, Y. Liu, Y.-S. Kim, A. R. Damadoran, Y. A. Huang, J. Xia, L. W. Martin, Y. Huang, J. A. Rogers, *Nat. Commun.* **2014**, *5*, 4496.

[8] C. Dagdeviren, S.-H. Hwang, Y. Su, S. Kim, H. Cheng, O. Gur, R. Haney, Y. Huang, J. A. Rogers, *Small* **2013**, *9*, 3398.

[9] P. M. Rørvik, T. Grande, M.-A. Einarsrud, *Adv. Mater.* **2011**, *23*, 4007.

[10] Y. Zhan, Y. Mei, L. Zheng, *J. Mater. Chem. C* **2014**, *2*, 1220.

[11] V. Maheshwari, R. Saraf, *Angew. Chem. Int. Ed.* **2008**, *47*, 7808.

[12] P. Martins, A. Lasheras, J. Gutierrez, J. M. Barandiaran, I. Orue, S. Lanceros-Mendez, S *J. Phys. D: Appl. Phys.* **2011**, *44*, 495303.

[13] M. Lee, C.-Y. Chen, S. Wang, S. N. Cha, Y. J. Park, J. M. Kim, L. J. Chou, Z. L. Wang, *Adv. Mater.* **2012**, *24*, 1759.

[14] D. S. Levi, N. Kusnezov, G. P. Carman, *Pediatr. Res.* **2008**, *63*, 552.
13

Published in Advanced Materials. Doi: [10.1002/adma.201506381](10.1002/adma.201506381) (2016).

[15] T. I. Kim, J. G. McCall, Y. H. Jung, X. Huang, E. R. Siuda, Y. Li, J. Song, J. M. Song, H. A. Pao, R.-H. Kim, C. Lu, S. D. Lee, I.-S. Song, G. Shin, R. Al-Hasani, S. Kim, M. P. Tan, J. Huang, F. G. Omenetto, J. A. Rogers, M. R. Bruchas, *Science* **2013**, *340*, 211.

[16] S. Egusa, Z. Wang, N. Chocat, Z. M. Ruff, A. M. Stolyarov, D. Shemuly, F. Sorin, P. T. Rakich, J. D. Jannopoulos, Y. Fink, *Nat. Mater.* **2010**, *9*, 643.

[17] Y. Qi, J. Kim, T. D. Nguyen, B. Lisko, P. Purohit, M. C. McAlpine, *Nano Lett.* **2011**, *11*, 1331.

[18] S. M. Nakhmanson, M. Buongiorno Nardelli, J. Bernholc, *Phys. Rev. Lett.* **2004**, *92*, 115504.

[19] T. Furukawa, *IEEE Trans. Electr. Insul.* **1989**, *24*, 375.

[20] S. –H. Bae, O. Kahya, B. K. Sharma, J. Kwon, H. J. Choc, B. Özyilmaz, J.-H. Ahn, *ACS Nano* **2013**, *7*, 3130.

[21] C. Hou, T. Huang, H. Wang, H. Yu, Q. Zhang Y. Li, *Sci. Rep.* **2013**, *3*, 3138.

[22] J. Fang, X. Wang, T. Lin, *J. Mater. Chem.* **2011**, *21*, 11088.

[23] K. Asadi, D. M. de Leeuw, B. de Boer, P. W. Blom, *Nat. Mater.* **2008**, *7*, 547.

[24] L. Persano, C. Dagdeviren, Y. Su, Y. Zhang, S. Girardo, D. Pisignano, Y. Huang, J. A. Rogers, *Nat. Commun.* **2013**, *4*, 1633.

[25] V. S. Bystrov, E. V. Paramonova, I. K. Bdikin, A. V. Bystrova, R. C. Pullar, A. L. Kholkin, *J. Mol. Model* **2013**, *19*, 3591.

[26] C. Chang, V. H. Tran, J. Wang, Y. K. Fuh, L. Lin, *Nano Lett.* **2010**, *10*, 726.

[27] R. Yang, Y. Qin, L. Dai, Z. L. Wang, *Nat. Nanotechnol.* **2009**, *4*, 34.

[28] J. Song, J. Zhou, Z. L. Wang, *Nano Lett.,* **2006**, *6*, 1656.

[29] X. Chen, S. Xu, N. Yao, W. Xu, Y. Shi, *Appl. Phys. Lett.* **2009**, *94*, 253113.

[30] S. A. Harfenist, S. D. Cambron, E. W. Nelson, S. M. Berry, A. W. Isham, M. M. Crain, K. M. Walsh, R. S. Keynton, R. W. Cohn, *Nano Lett.* **2004**, *4*, 1931.
14

Figures and Tables

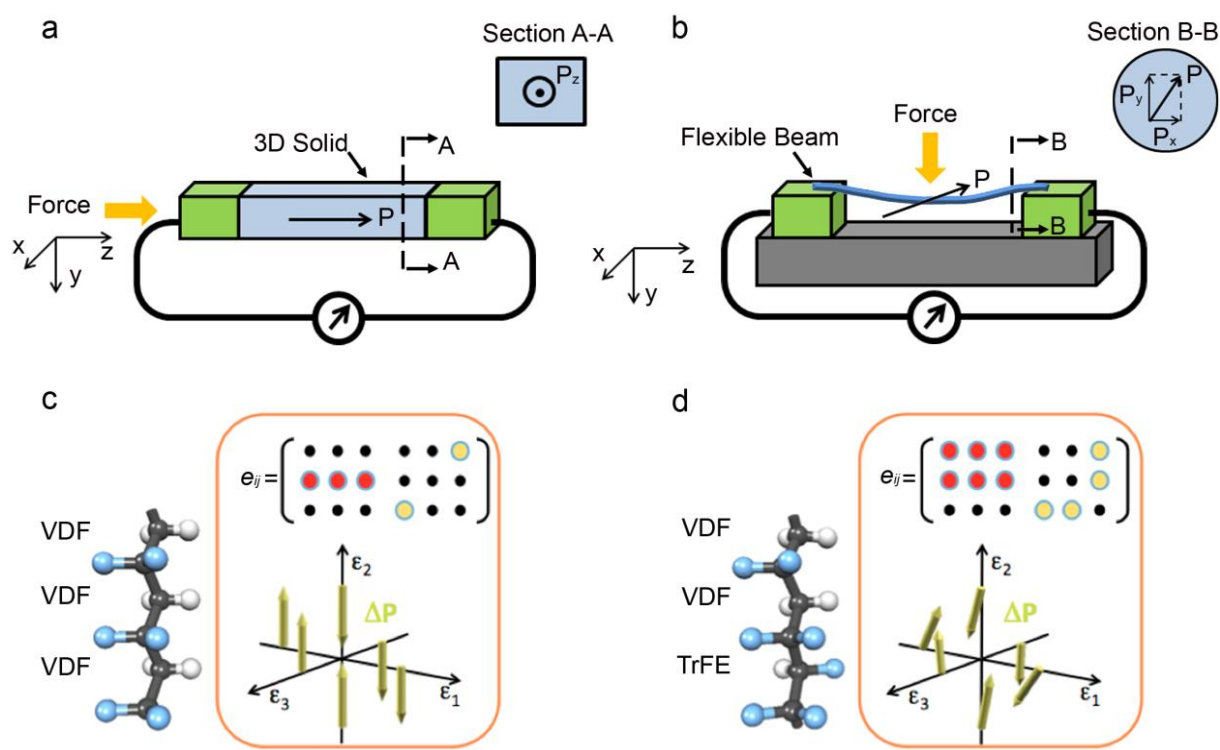

**Figure 1.** Schematics of typical uniaxial piezoelectricity obtained through normal strain on a 3D crystalline solid (a), and of our experimental configuration with piezoelectric P(VDF-TrFE) nanobeams (b), together with the corresponding cross-sectional schemes highlighting the polarization components for a generic uniaxial solid (rectangle in the top-right inset, a) and for a biaxial fiber (circle, b). The used (*x*, *y* and *z*) coordinates are also shown in the schemes. (c, d) Side view of VDF (c) and VDF-TrFE (d) structures, resulting from total-energy-and forces optimization (see the text). Dark grey, white and cyan label respectively C, H and F atoms. Green arrows define the spatial direction of polarization $P_{\varepsilon=0}$ for the minimum energy unstrained systems. Inset boxes show the corresponding sketches of piezoelectric tensors ($e_{ij}$). Small (large) dots in the $e_{ij}$ matrix plot correspond to zero (non-zero) components, whereas red (yellow) color corresponds to normal (shear) strain, respectively.

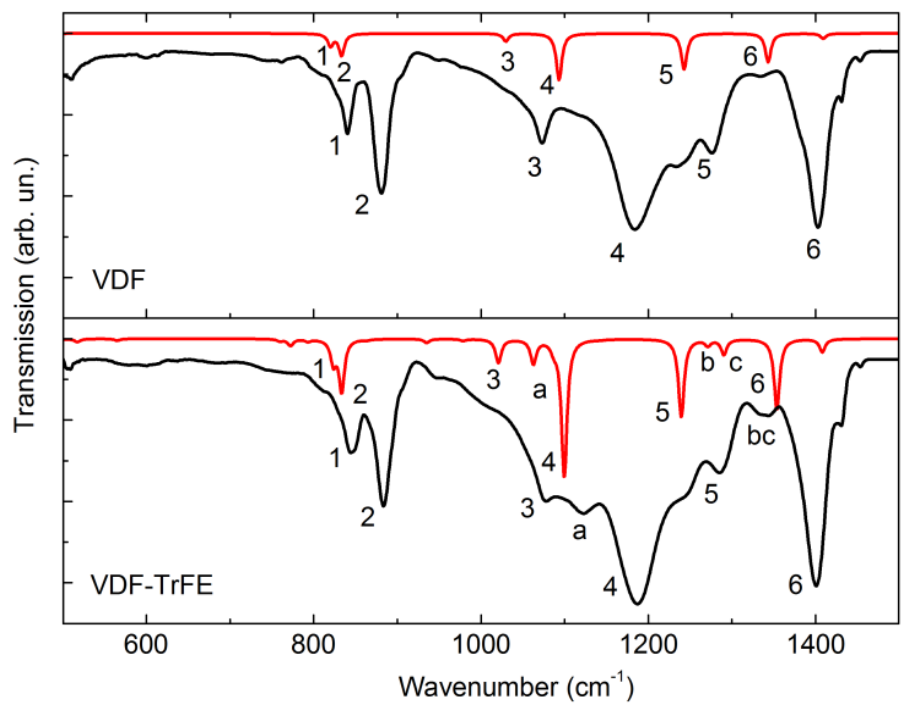

**Figure 2.** FTIR transmission spectra. Comparison of experimental (black lines) and theoretical (red lines) spectra for VDF and VDF-TrFE compounds. Numbers (1-6) identify spectroscopic features common to both systems (~845 cm$^{-1}$, ~885 cm$^{-1}$, ~1074 cm$^{-1}$, ~1186 cm$^{-1}$, ~1280 cm$^{-1}$ and ~1402 cm$^{-1}$), letters (a-c) mark vibrational peaks (1065 cm$^{-1}$, 1270 cm$^{-1}$, and 1287 cm$^{-1}$) associated to combined C-F and C-H rocking modes of the TrFE units. Spectra are vertically shifted for better clarity.

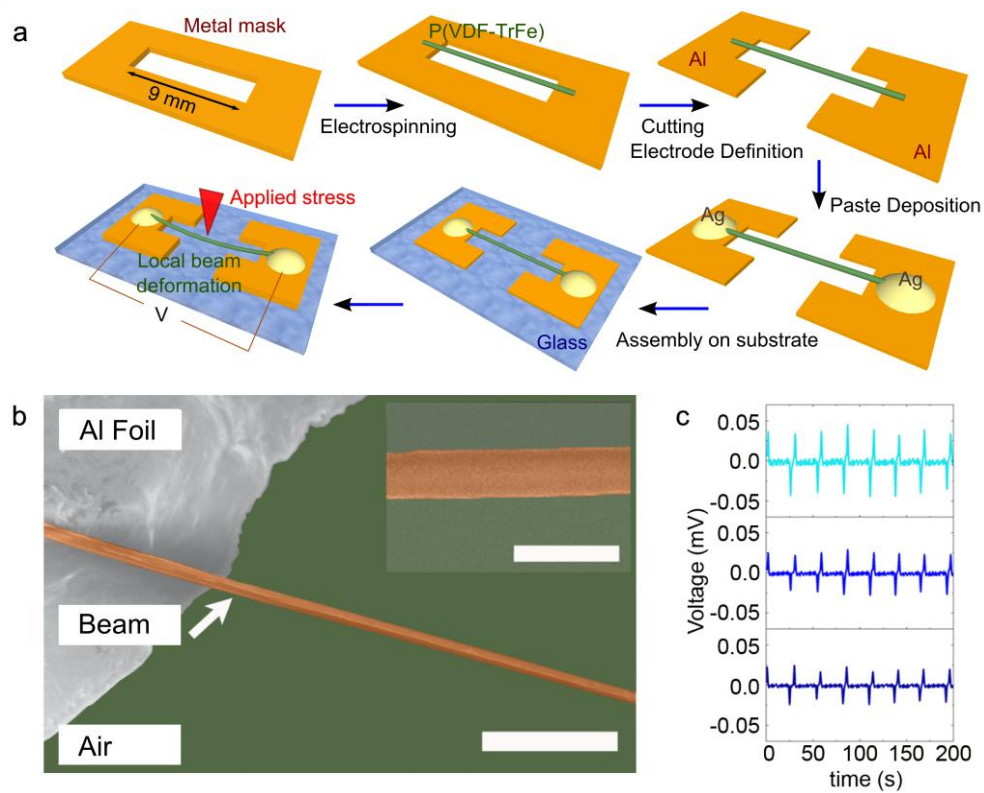

**Figure 3.** (a) Process schematics illustrating the fabrication steps for piezoelectric devices based on suspended P(VDF-TrFE) nanobeams. Features not in scale. (b) SEM micrograph of a suspended piezoelectric wire with the edge lying on metal (scale bar, 5 μm). Inset: High magnification micrograph of the wire (scale bar, 1 μm). (c) Measured output voltage under repeated load/unload cycles for the suspended wire. From bottom to top: vertical wire displacements are 38 nm, 48 nm, 109 nm.

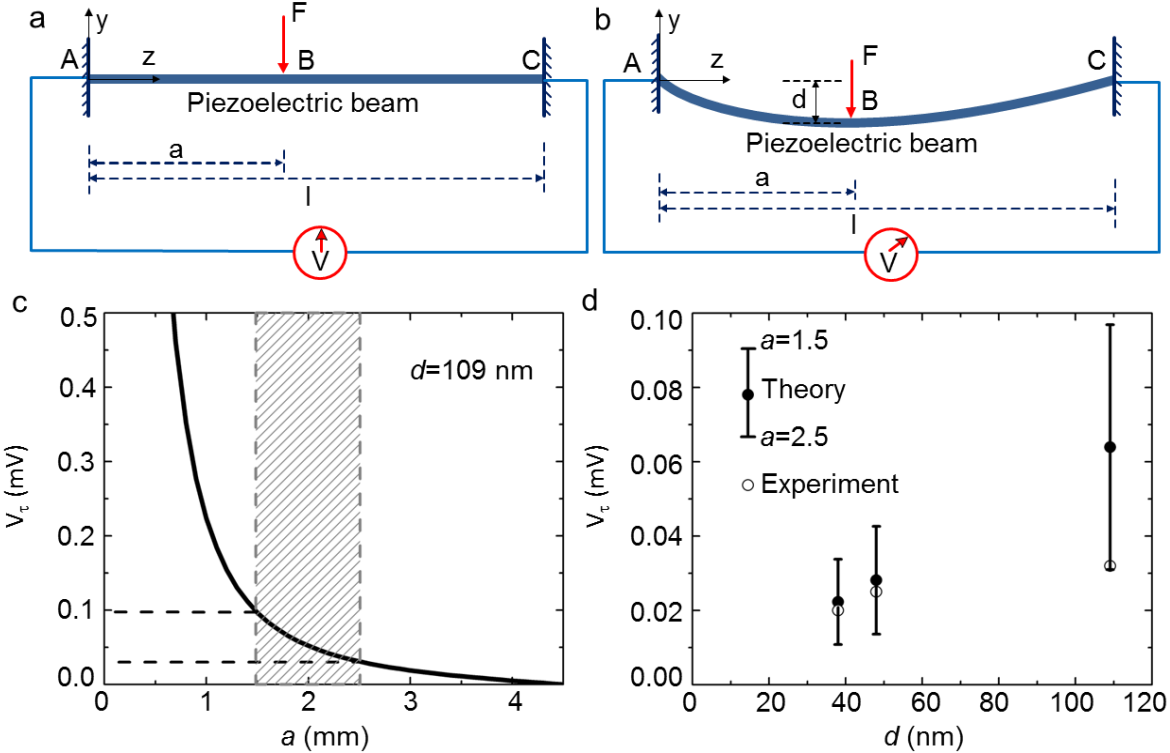

**Figure 4.** Illustration of the analytic model for the response of a suspended piezoelectric nanobeam under concentrated eccentric loading: (a) undeformed, (b) deformed. (c) Theoretical transverse piezo-electric voltage response generated by shear stress at different positions of the bending point with $d = 109$ nm. (d) Comparison between experiment and theoretical transverse piezo-electric voltage response generated by shear stress ($a$=1.5-2.5mm).

Table 1. Calculated, normal components of the proper piezoelectric tensor and of the diagonal elements of the dielectric tensor for PVDF and P(VDF-TrFE) (VDF:TrFE ratio ≅ 70:30). The $e_{11}$ range refers to variations calculated for diluted and clustered systems (see the main text). No variation is found for other components.

|  | **PVDF** | **P(VDF-TrFE)** |
|---|---|---|
| $e_{11}$ (C/m$^2$) | 0 | -0.14 - -0.05 |
| $e_{22}$ (C/m$^2$) | -0.32 | -0.28 |
| $e_{33}$ (C/m$^2$) | 0 | 0 |
| $\tilde{k}_{11}$ | 2.16 | 2.18 |
| $\tilde{k}_{22}$ | 2.14 | 2.17 |
| $\tilde{k}_{33}$ | 2.36 | 2.36 |

Table 2. Calculated shear components of the proper piezoelectric tensor for PVDF and P(VDF-TrFE) (VDF:TrFE ratio ≅ 70:30). $e_{34}$ range: calculated variations between diluted and clustered systems (see the main text). No variation is found for other components.

|  | **PVDF** | **P(VDF-TrFE)** |
|---|---|---|
| $e_{34}$ (C/m$^2$) | 0.13 | 0.11-0.12 |
| $e_{35}$ (C/m$^2$) | 0 | 0.01 |
| $e_{36}$ (C/m$^2$) | 0 | 0 |

# Supporting Information

**Shear piezoelectricity in poly(vinylidenefluoride-*co*-trifluoroethylene): full piezotensor coefficients by molecular modeling, biaxial transverse response, and use in suspended energy-harvesting nanostructures**

*Luana Persano*, *Alessandra Catellani*, *Canan Dagdeviren*, *Yinji Ma*, *Xiaogang Guo*, *Yonggang Huang*, *Arrigo Calzolari*, and *Dario Pisignano*

Dr. L. Persano, Prof. D. Pisignano
CNR-NANO, Istituto Nanoscienze, Euromediterranean Center for Nanomaterial Modelling and Technology (ECMT), via Arnesano I-73100, Lecce (Italy)
E-mail: luana.persano@nano.cnr.it
Dr. A. Catellani and Dr. A. Calzolari
CNR-NANO, Istituto Nanoscienze, Centro S3, via Campi 213, I-41125 Modena (Italy)
Dr. C. Dagdeviren
Department of Materials Science and Engineering, Frederick Seitz Materials Research Laboratory, and Beckman Institute for Advanced Science, University of Illinois at Urbana-Champaign, Urbana, IL 61801 (USA).
The David H. Koch Institute for Integrative Cancer Research, Massachusetts Institute of Technology, Cambridge, MA 02139 (USA)
Harvard Society of Fellows, Harvard University, Cambridge, MA 02138, USA
Dr. Y. Ma, Dr. X. Guo and Prof. Y. Huang
Department of Civil and Environmental Engineering and Department of Mechanical Engineering, Northwestern University, Evanston, IL 60208 (USA)
Dr. Y. Ma
Department of Engineering Mechanics, Center for Mechanics and Materials, Tsinghua University, Beijing 100084 (China)
Dr. X. Guo
College of Aerospace and Civil Engineering, Harbin Engineering University, Harbin 150001 (China)
Prof. D. Pisignano
Dipartimento di Matematica e Fisica "Ennio De Giorgi", Università del Salento, via Arnesano I-73100 Lecce (Italy)

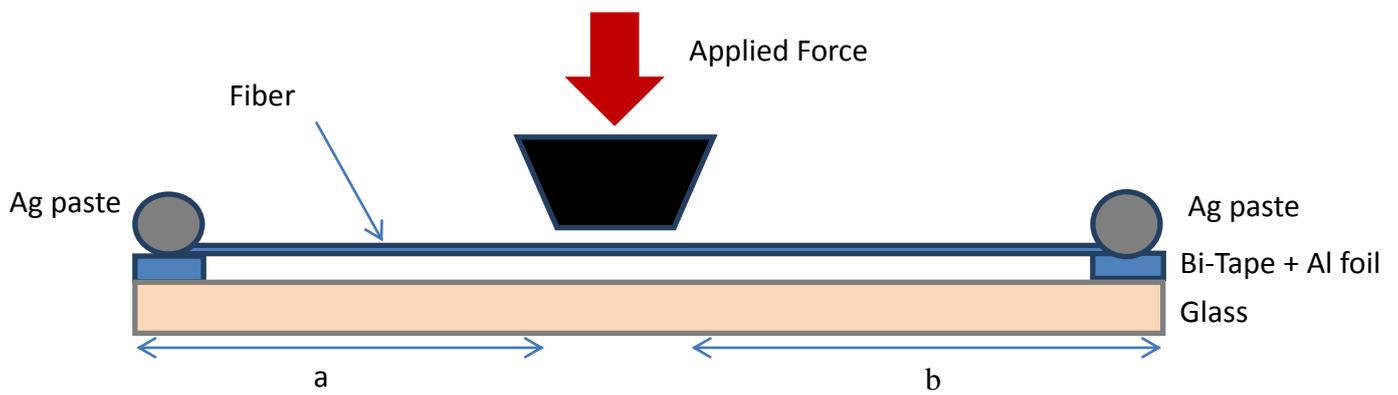

**Figure S1.** Force-Indentation set-up. Schematic illustration of the experimental setup for force-indentation measurements on single suspended fibers.

**Note 1. Supplemental theoretical results**

For each system we performed a full atomic relaxation of both the lattice parameters and the atomic positions in the cell, towards the minimum energy configuration. We evaluated the polarization vector (**P**), and the dielectric tensor for each crystalline assembly, upon the application of normal and shear strain along the three spatial directions. All numerical data are reported in the following. We also report the most relevant Born charges ($Z^*$) of the studied systems, which express the variation of the polarization response due to an atomic displacement. For P(VDF-TrFE) systems, the copolymers were simulated with a VDF:TrFE ratio matching as closely as possible that measured experimentally (70:30), and optimizing the computational effort. Indeed, for the diluted system discussed below, a complete information could be obtained by exploiting the periodicity of the system and consequently reducing the number of atoms per cell.

In the pictures below, we report the side view of the single component chains (left) and their arrangement in the crystalline bulk (right). The inequivalent atomic species and the direction of the polarization vectors are also reported, as well as the resulting geometric data, such as the optimized cell dimensions, the atomic distances and angles.

VDF polymer

The VDF and VDF-TrFE (diluted) crystals include 2 chains per cell and three monomers per chain. In the former case only VDF species are considered; in the latter 2 VDF units and 1 TrFE monomer define the structures. Due to the periodic boundary conditions, no two consecutive TrFE units occur along the chain, thus keeping the copolymers dilutely distributed. The comparison between the two systems highlights that the inclusion of TrFE imparts local distortions along the chain reducing the ideal order of the pristine VDF case.

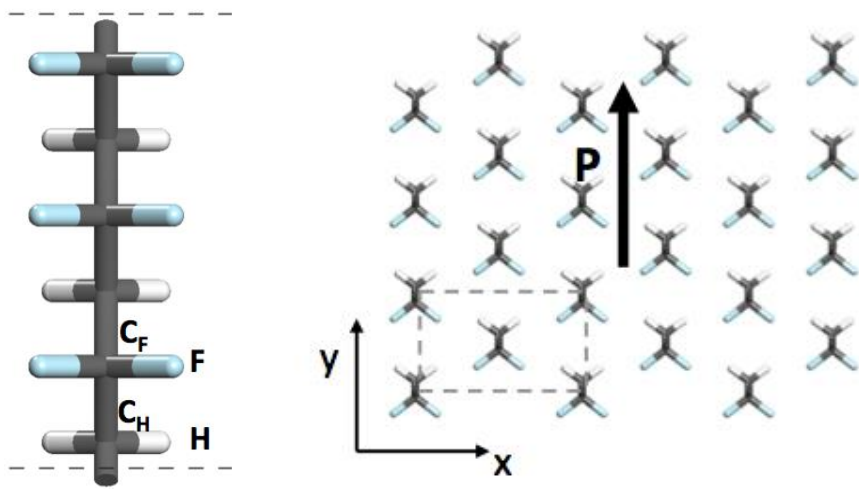

Lattice parameters: $(8.59 \times 4.91 \times 7.78)$ Å$^3$

Average distances:

$C_H$-$C_F$ = 1.54 Å, $C_H$-H = 1.10 Å, $C_F$-F = 1.39 Å

Average angles: (H-$C_H$-H) = 108.6°, (F-$C_F$-F) = 106.9°, ($C_H$-$C_F$-$C_H$) = 114.5°

Total polarization: $\boldsymbol{P} \equiv (0.00, 0.20, 0.00)$ C/m$^2$, $|\boldsymbol{P}| = 0.20$ C/m$^2$

Dielectric tensor in cartesian axis $\epsilon = \begin{pmatrix} 2.16 & 0.00 & 0.00 \\ 0.00 & 2.14 & 0.00 \\ 0.00 & 0.00 & 2.36 \end{pmatrix}$

Born charges $\left(Z^* = -\Omega \frac{\partial^2 E}{\partial r \partial \mathbb{E}}\right)$, in units of elemental charge $e$: $C_H$ = -0.28, $C_F$ = +1.49, H = +0.11, F = -0.82

Proper piezoelectric tensor (in units C/m$^2$): $e_{ij} = \begin{pmatrix} 0 & 0 & 0 & 0 & 0 & +0.10 \\ -0.11 & -0.32 & -0.11 & 0 & 0 & 0 \\ 0 & 0 & 0 & +0.13 & 0 & 0 \end{pmatrix}$

VDF:TrFE – *Diluted* structure

As discussed in the main text, the inclusion of the TrFE units, imparting a rotation of the F-C-F units in the *xy* plane, leads to the misalignment of the polarization vector and to a slight reduction of its module. On the other hand, the presence of polarization components in the *xy* plane leads to the presence of not-null components in the normal and shear piezoelectric tensor (see main text).

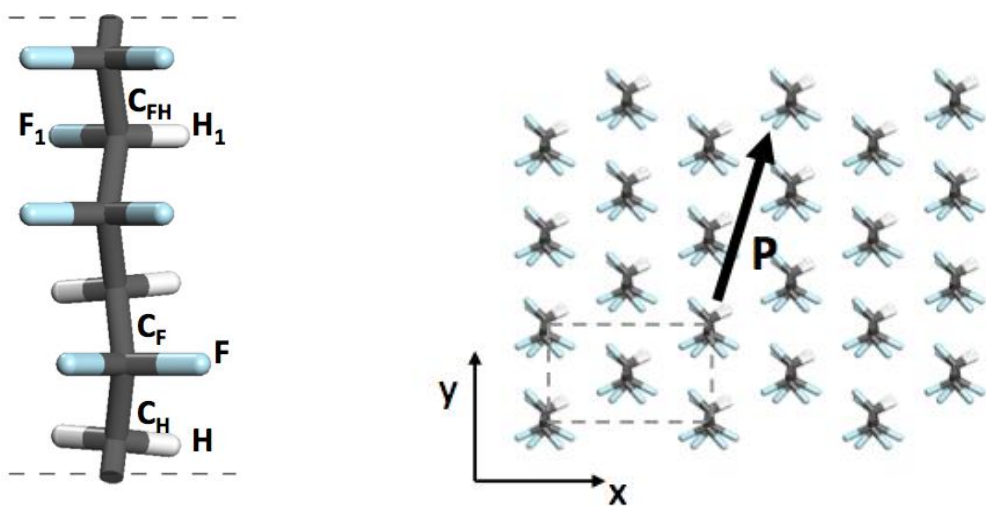

Lattice parameters: (8.68 × 4.87 × 7.73) Å$^3$

Average distances:

$C_H$-$C_F$ = 1.53 Å, $C_H$-H = 1.10 Å, $C_F$-F = 1.40 Å,

$C_F$-$C_{FH}$ = 1.55 Å, $C_{FH}$-$H_1$ = 1.10 Å, $C_{FH}$-$F_1$ = 1.41 Å

Average angles:

($C_H$- $C_F$-$C_H$) = 114.6°, ($C_F$- $C_H$- $C_F$) = 113.8°, ($C_F$- $C_{FH}$-$C_F$) = 116.6°

(H-$C_H$-H) = 107.8°, (F- $C_F$- F) = 104.7°, ($H_1$-$C_{FH}$-$F_1$) = 108.8°

Total polarization:

$\boldsymbol{P} \equiv (0.05, 0.15, 0.00)$ C/m$^2$,  $|\boldsymbol{P}| = 0.16$ C/m$^2$

Dielectric tensor in cartesian axis $\epsilon = \begin{pmatrix} 2.18 & -0.11 & 0.00 \\ -0.11 & 2.17 & 0.00 \\ 0.00 & 0.00 & 2.36 \end{pmatrix}$

Born charges $\left(Z^* = -\Omega \frac{\partial^2 E}{\partial r \partial \mathbb{E}}\right)$ in units of elemental charge *e*:

$C_H$ = -0.30, $C_F$ = +1.43, $C_{FH}$ = +0.42, H = +0.04, $H_1$ = 0.11, F = -0.82, $F_1$ = -0.68

Proper piezoelectric tensor (in units C/m$^2$):   $e_{ij} = \begin{pmatrix} -0.14 & -0.03 & +0.05 & 0 & 0 & -0.01 \\ -0.08 & -0.28 & -0.07 & 0 & 0 & +0.04 \\ 0 & 0 & 0 & +0.11 & +0.01 & 0 \end{pmatrix}$

VDF:TrFE – *Clustered* structure

Clustered systems are simulated including two parallel VDF-TrFE chains in the simulation cell. Each chain is composed of 7 VDF units and 3 consecutive TrFE units. In order to obtain clustered consecutive TrFE units along the chain a larger cell (10 units per polymer) must be taken into account. After complete relaxation of both lattice parameters and internal nuclei coordinates, chains undergo structural distortions which include both bending of the C-backbone and in-plane rotation of the C-F bonds. We report in the following the main structural and polarization properties of the clustered system.

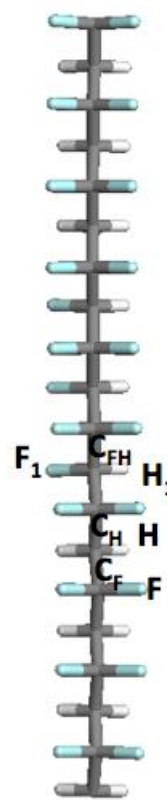

Lattice parameters: $(8.68 \times 4.87 \times 25.70)$ Å$^3$

Average distances:

$C_H$-$C_F$ = 1.53 Å, $C_H$-H = 1.10 Å, $C_F$-F = 1.40 Å,

$C_F$-$C_{FH}$ = 1.54 Å, $C_{FH}$-$H_1$ = 1.10 Å, $C_{FH}$-$F_1$ = 1.40 Å

Average angles:

($C_H$- $C_F$-$C_H$) = 114.6°, ($C_F$- $C_H$- $C_F$) = 113.1°, ($C_F$- $C_{FH}$-$C_F$) = 115.8°

(H-$C_H$-H) = 107.8°, (F- $C_F$- F) = 104.7°, ($H_1$-$C_{FH}$-$F_1$) = 108.9°

Total polarization: $\boldsymbol{P} \equiv (0.02, 0.01, 0.00)$ C/m$^2$, $|\boldsymbol{P}| = 0.02$ C/m$^2$

Dielectric tensor in cartesian axis $\epsilon = \begin{pmatrix} 2.18 & -0.11 & 0.00 \\ -0.11 & 2.17 & 0.00 \\ 0.00 & 0.00 & 2.36 \end{pmatrix}$

Born charges $\left(Z^* = -\Omega \frac{\partial^2 E}{\partial r \partial \mathbb{E}}\right)$, in units of elemental charge $e$:

$C_H$ = -0.30, $C_F$ = +1.43, $C_{FH}$ = +0.42, H = +0.04, $H_1$=0.11, F=-0.82, $F_1$=-0.68

Proper piezoelectric tensor (in units C/m$^2$): $e_{ij} = \begin{pmatrix} -0.05 & -0.01 & 0 & 0 & 0 & +0.09 \\ -0.04 & -0.28 & -0.01 & 0 & 0 & +0.02 \\ 0 & 0 & 0 & +0.12 & +0.01 & 0 \end{pmatrix}$

The polarization is not completely suppressed and it is still along the plain perpendicular to the chain axis, which may still give rise to an effective piezoelectric behavior as also suggested by the similar normal ($e_{22}$=-0.28 C/m$^2$) and shear piezoelectric coefficients ($e_{34}$=0.11-0.12 C/m$^2$ and $e_{35}$=0.01 C/m$^2$) obtained for the diluted and for the clustered polymers.

**Note 2. Electromechanical analysis of a piezoelectric fiber under an eccentric loading**

**2.1 Mechanics analysis**

A mechanics model of a beam with both ends clamped is established for the suspended piezoelectric fiber, and the beam is subjected by a concentrated, eccentric loading $F$ (Figs. 4a and b). The model gives analytically the reaction forces $F_A$ and $F_C$ and moments $M_A$ and $M_C$ at the left and right clamps as:

$$F_A = \frac{F(l-a)^2(2a+l)}{l^3}, \tag{S1a}$$

$$F_C = \frac{Fa^2(3l-2a)}{l^3}, \tag{S1b}$$

$$M_A = \frac{Fa(l-a)^2}{l^2}, \tag{S1c}$$

$$M_C = \frac{Fa^2(l-a)}{l^2}. \tag{S1d}$$

The deflection of piezoelectric fiber is

$$w = \begin{cases} -\dfrac{2M_A}{\pi E r^4} z^2 + \dfrac{2F_A}{3\pi E r^4} z^3 & (0 \leq z \leq a) \\ -\dfrac{2M_A}{\pi E r^4} z^2 + \dfrac{2F_A}{3\pi E r^4} z^3 - \dfrac{2F}{3\pi E r^4}(z-a)^3 & (a < z \leq l) \end{cases}, \tag{S2}$$

where $r$ and $E$ are the fiber radius and elastic modulus, respectively, and the coordinate $z$ is along the fiber length (Fig. 4a). The above equation can also be expressed in terms of the displacement $d$ at the loading point as:

$$w = \begin{cases} -\dfrac{3ld}{2a^2(l-a)} z^2 + \dfrac{(2a+l)d}{2a^3(l-a)} z^3 & (0 \leq z \leq a) \\ -\dfrac{3ld}{2a^2(l-a)} z^2 + \dfrac{(2a+l)d}{2a^3(l-a)} z^3 - \dfrac{l^3 d}{2a^3(l-a)^3}(z-a)^3 & (a < z \leq l) \end{cases}. \quad (S3)$$

The normal (membrane) strain ($\varepsilon_{33}$) caused by the stretching force is defined as the elongation of the nanobeam under the application of a concentrated eccentric loading, $F$ (Fig. S2):

$$\varepsilon_{33} = \frac{(dz' - dz)}{dz} = \frac{\left(\sqrt{(dz)^2 + (dw)^2} - dz\right)}{dz} \approx \frac{1}{2}\left(\frac{dw}{dz}\right)^2 \quad (S4)$$

The shear strain in the piezoelectric fiber is:

$$\varepsilon_{23} = \begin{cases} \dfrac{3Er^2 d(2a+l)}{4a^3(l-a)G} & (0 \leq z \leq a) \\ -\dfrac{3Er^2 d(3l-2a)}{4a(l-a)^3 G} & (a < z \leq l) \end{cases}, \quad (S5)$$

where $G$ is the shear modulus.

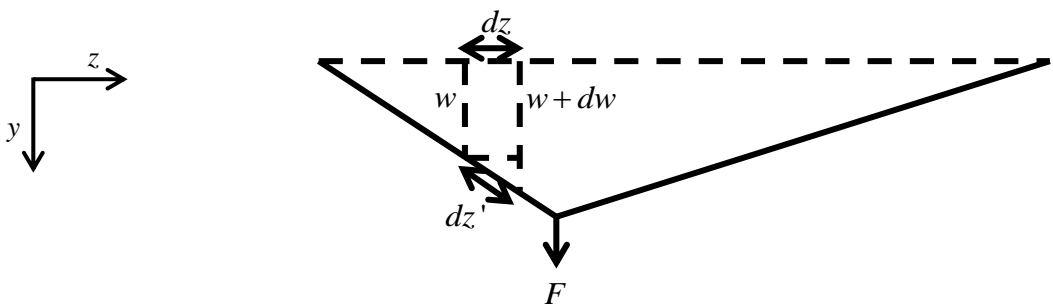

**Figure S2.** Schematic illustration of the elongation of a suspended piezoelectric nanobeam under the application of a concentrated eccentric loading, $F$.

**2.2 Voltage**

The constitutive model of piezoelectric materials relates the piezoelectric tensor $e_{ij}$, the strain $\varepsilon_{ij}$, the dielectric tensor $k_{ij}$, the electric field $E_i$ and the electric displacement $D_i$ as:

$$\begin{Bmatrix} D_1 \\ D_2 \\ D_3 \end{Bmatrix} = \begin{Bmatrix} e_{11} & e_{12} & e_{13} & e_{14} & e_{15} & e_{16} \\ e_{21} & e_{22} & e_{23} & e_{24} & e_{25} & e_{26} \\ e_{31} & e_{32} & e_{33} & e_{34} & e_{35} & e_{36} \end{Bmatrix} \begin{Bmatrix} \varepsilon_{11} \\ \varepsilon_{22} \\ \varepsilon_{33} \\ 2\varepsilon_{23} \\ 2\varepsilon_{31} \\ 2\varepsilon_{12} \end{Bmatrix} + \begin{Bmatrix} k_{11} & 0 & 0 \\ 0 & k_{22} & 0 \\ 0 & 0 & k_{33} \end{Bmatrix} \begin{Bmatrix} E_1 \\ E_2 \\ E_3 \end{Bmatrix} \quad (S6)$$

For $\varepsilon_{11} = \varepsilon_{22} = \varepsilon_{12} = \varepsilon_{13} = 0$, Eq. (S6) gives:

$$D_3 = e_{33}\varepsilon_{33} + 2e_{34}\varepsilon_{23} + k_{33}E_3 . \quad (S7)$$

Let $V = \int_0^l E_3 dz$ denote the voltage bias along the fiber length, which gives:

$$\int_0^l D_3 dz = e_{33}\int_0^l \varepsilon_{33} dz + 2e_{34}\int_0^l \varepsilon_{23} dz + k_{33}V . \quad (S8)$$

The voltage $V$ and the current $I = -\pi r^2 \dfrac{\partial \int_0^l D_3 dz}{l \partial t}$ are related by the resistance $R$ of the voltmeter, $V=IR$, which, together with Eq. (S8), gives:

$$\frac{dV}{dt} + \frac{l}{\pi r^2 R k_{33}} V = -\frac{e_{33}\partial \int_0^l \varepsilon_{33} dz}{k_{33} \partial t} - \frac{2e_{34}\partial \int_0^l \varepsilon_{23} dz}{k_{33} \partial t} . \quad (S9)$$

For a step-function load, Eq. (S9) becomes:

$$\frac{dV}{dt} + \frac{l}{\pi r^2 R k_{33}} V = -\frac{e_{33}\int_0^l \varepsilon_{33}dz + 2e_{34}\int_0^l \varepsilon_{23}dz}{k_{33}} \delta(t), \quad \text{(S10)}$$

where $\delta(t)$ is a Delta function. For the initial condition $V(0)=0$, Eq. (S10) has the solution:

$$V(t) = -\frac{e_{33}\int_0^l \varepsilon_{33}dz + 2e_{34}\int_0^l \varepsilon_{23}dz}{k_{33}} \exp\left(-\frac{lt}{\pi r^2 R k_{33}}\right), \quad \text{(S11)}$$

which has the maximum value, $V_{max} = -\frac{e_{33}\int_0^l \varepsilon_{33}dz + 2e_{34}\int_0^l \varepsilon_{23}dz}{k_{33}}$. Here, $-\frac{2e_{34}\int_0^l \varepsilon_{23}dz}{k_{33}}$ represents the transverse piezo-electric voltage $V_\tau$ (contribution from the shear strain), which leads to Eq. 1, and $-\frac{e_{33}\int_0^l \varepsilon_{33}dz}{k_{33}}$ represents the normal piezo-electric voltage $V_\sigma$ (contribution from the normal strain) which is given by:

$$V_\sigma = -\frac{3e_{33}l^3}{20k_{33}a^2(l-a)^2} d^2. \quad \text{(S12)}$$

Finally, $I=V/R$, which, together with Eq. (S11), gives:

$$I(t) = -\frac{e_{33}\int_0^l \varepsilon_{33}dz + 2e_{34}\int_0^l \varepsilon_{23}dz}{Rk_{33}} \exp\left(-\frac{lt}{\pi r^2 R k_{33}}\right). \quad \text{(S13)}$$

This involves the maximum transverse piezoelectric current, $I_\tau$ (contribution from the shear strain), and the maximum normal piezoelectric current, $I_\sigma$ (contribution from the normal strain), which are given by:

$$I_\tau = \frac{3e_{34}r^2l(2a-l)(1+v)}{a^2(l-a)^2 k_{33} R} d. \quad \text{(S14a)}$$

$$I_\sigma = -\frac{3e_{33}l^3}{20a^2(l-a)^2 k_{33} R} d^2 .$$
(S14b)

The maximum transverse piezo-electric voltage ($V_\tau$) and its dependence on the displacement, $d$, are shown in Fig. 4c and in Fig. S3a, respectively. For $e_{33} = 0$, one obtains zero normal piezoelectric voltage. The current resulting from using a small (1 kΩ) load resistance is shown in Fig. S3b. Instead, assuming a non-zero value of $e_{33}$ (e.g. $e_{33}$ = $10^{-5}$, $10^{-4}$, or $10^{-3}$ C/m$^2$), the normal piezo-electric voltage ($V_\sigma$) will have the behavior shown in Fig. S4. For $e_{33}$ values >$10^{-4}$, $V_\sigma$ becomes significantly larger than experimentally measured piezo-voltages as shown in Fig. S4a, which sets an upper limit for the normal piezo-coefficient determining the response to stretching in the real polymeric nanowire.

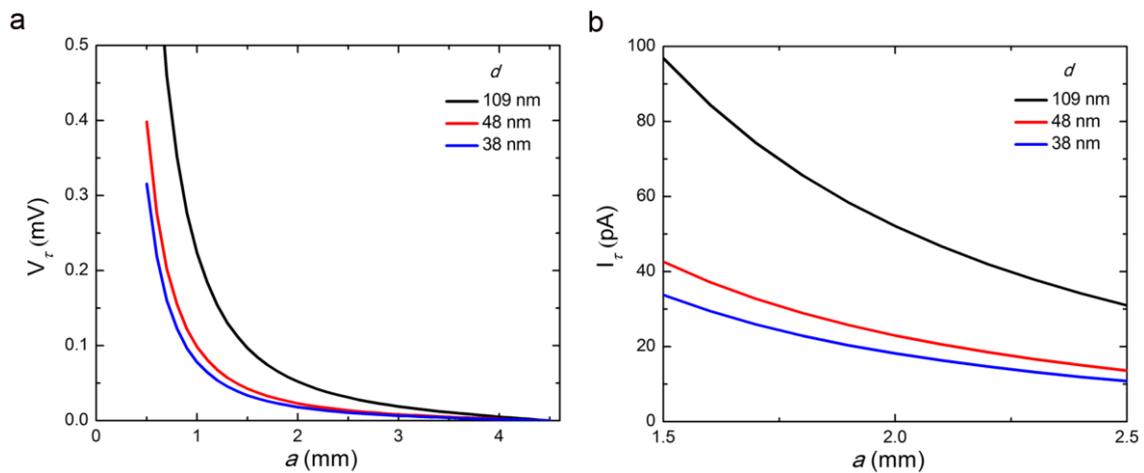

**Figure S3**. (a) Theoretical, maximum $V_\tau$ response at different positions of the bending point along the fiber longitudinal axis ($a$) for three vertical displacement values, 38 nm, 48 nm and 109 nm. (b) Maximum corresponding currents for $R$= 1 kΩ and for 1.5 mm ≤ $a$ ≤ 2.5 mm, as indicated in Fig. 4c,d.

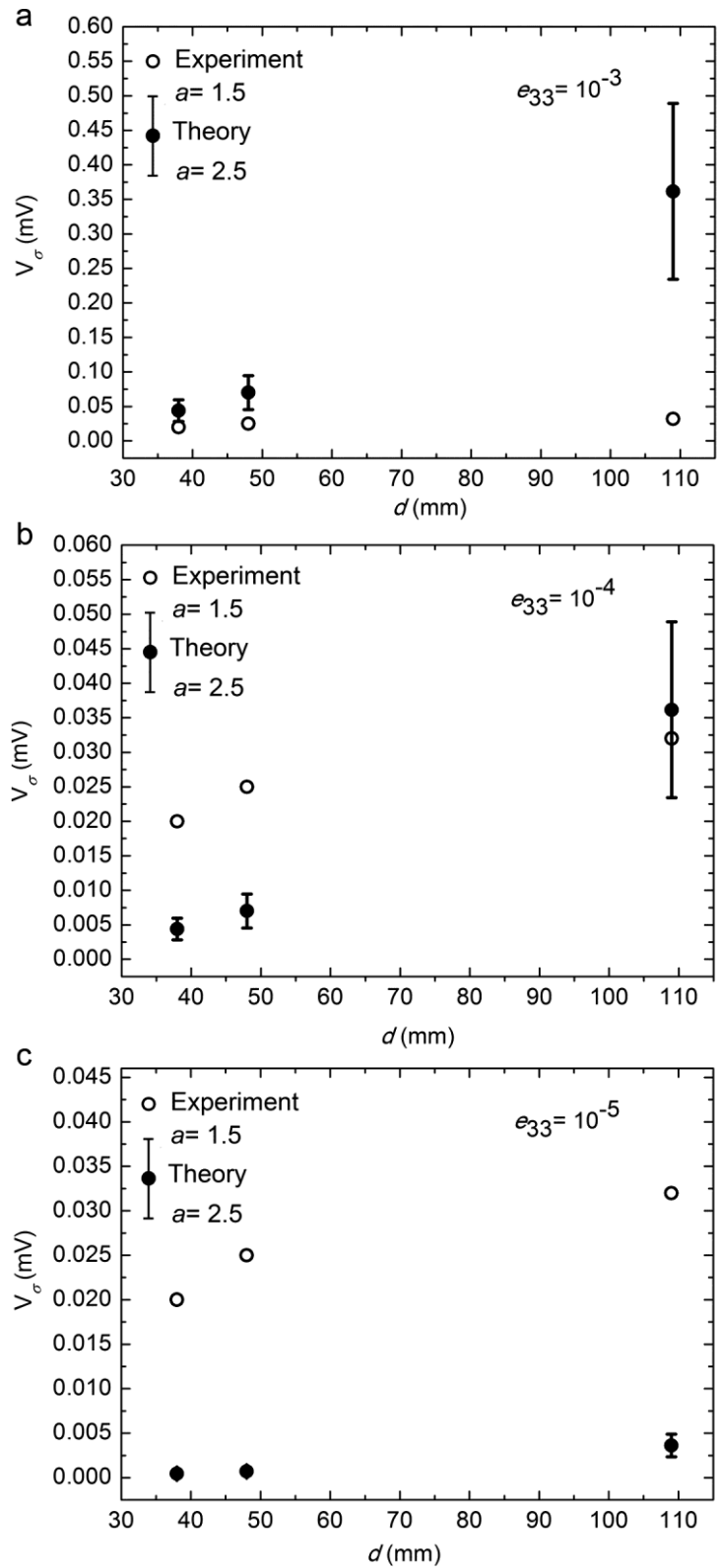

**Figure S4.** Comparison between experiment and theoretical normal piezo-electric voltage response generated by stretching stress ($a$=1.5-2.5mm). $V_\sigma$ is obtained by considering $e_{33} = 10^{-3}$ C/m$^2$ (a), $10^{-4}$ C/m$^2$ (b), and $10^{-5}$ C/m$^2$ (c), respectively.

**Note 3. Analytical definitions**

Despite its intuitive meaning, the theoretical description of spontaneous polarization from an atomistic point of view is a subtle problem, solved throughout a theoretical approach based on the evaluation of the Berry phase (BP) of the electronic structure.[S1] The polarization difference between two states (α, β) of a system is computed as a geometrical quantum phase $\Delta \boldsymbol{P} = \boldsymbol{P}^\alpha - \boldsymbol{P}^\beta$. The polarization vector for each state $\boldsymbol{P}^\alpha$ can be split into two parts $\boldsymbol{P}^\alpha_{ion}$, $\boldsymbol{P}^\alpha_{el}$ corresponding to the ionic and electronic contributions, respectively. For spin restricted systems, the expression for the total polarization can be written as follow:

$$\boldsymbol{P}^\alpha = \boldsymbol{P}^\alpha_{ion} + \boldsymbol{P}^\alpha_{el} = \frac{e}{\Omega}\sum_\tau Z^\alpha_\tau \mathbf{r}^\alpha_\tau - \frac{2ie}{8\pi^3} \sum_n \int_{BZ} d\boldsymbol{k}\ \langle u^\alpha_{nm}|\boldsymbol{\nabla}_m|u^\alpha_{nm}\rangle, \qquad (S15)$$

where $\Omega$ is the volume of the unit cell, $Z_\tau$ and $\mathbf{r}_\tau$ are the charge and position of the τ-th atom in the cell, and $u_{nm}$ are periodic parts of the occupied Bloch states of the system. For the electronic part, an electronic phase $\varphi^\alpha_i$ (BP) defined modulo $2\pi$ can be introduced as:

$$\varphi^\alpha_i = \Omega \mathbf{G}_i \cdot \boldsymbol{P}^\alpha_{el} / e\ , \qquad (S16)$$

where $\mathbf{G}_i$ is the reciprocal lattice vector in the $i$ direction. For a bulk system the spontaneous polarization is a multivalued quantity, i.e. "branch" dependent, defined modulo $\frac{e}{\Omega}\boldsymbol{R}$, where $\boldsymbol{R}$ is a vector of the three-dimensional direct lattice. The *direct* piezoelectric tensor is defined as derivative of the spontaneous polarization ($\boldsymbol{P}$) with respect to an applied strain ($\varepsilon$), at finite (zero) external electric field ($\mathbb{E}$):

$$\widetilde{e_{ilk}} = \left.\frac{\partial P_i}{\partial \varepsilon_{lk}}\right|_{\mathbb{E}=0}\ , \qquad (S17)$$

where indexes $i, l, k = 1, 2, 3$ run over the three spatial directions. The coefficients $\widetilde{e_{ilk}}$ are usually known as the *improper* piezoelectric coefficients, related only to the static polarization term of an isolated bulk sample under strain (Figure S5a). As the spontaneous polarization is a multivalued quantity, the improper coefficients result to be branches dependent.

In standard experiments, however, the piezoelectric material is not electrically isolated, but rather connected to an electrical circuit (Figure S5b). The so-called *proper* piezoelectric tensor $e_{ilk}$ includes the depolarizing fields and the polarization-induced interface charges. *Proper* piezoelectric coefficients are those to be compared with experiments and are independent of polarization branches. The *proper* tensor can be evaluated starting from the *improper* one:[S2]

$$e_{ilk} = \widetilde{e_{ilk}} + \delta_{lk}P_i - \delta_{il}P_k. \tag{S18}$$

The piezoelectric response can be decomposed into a sum of two parts: a "clamped-ion" part and an "internal" strain part. The former consists of a homogeneous strain in which nuclear coordinates are not allowed to relax, the latter describes the internal distortion of nuclei induced by fixed strain. Clamped-ion terms are generally not physically relevant, and the more subtle internal relaxation has to be taken into account for a direct comparison with experiments. Numerical values reported in the main text were computed fully relaxing the atomic structure of the systems for each external strain applied to the simulation cell.

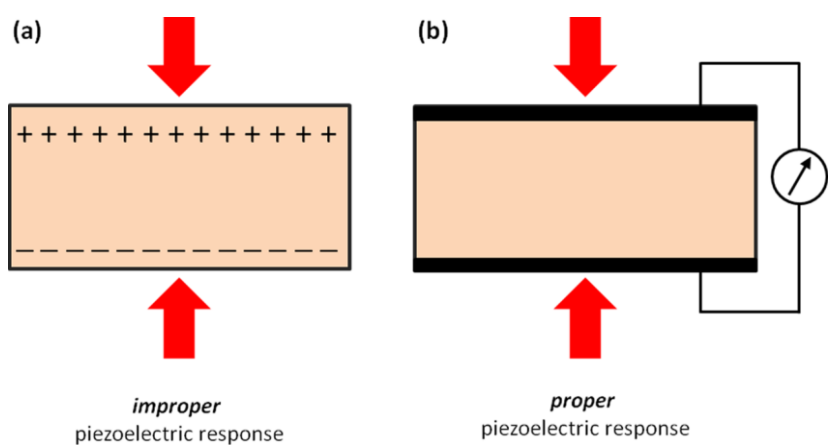

**Figure S5.** Illustrative schemes of experimental setup for the evaluation of (a) improper and (b) proper piezoelectric coefficients. Strain ε (red arrows) is applied to a generic piezoelectric material (shaded area).

**Note 4. Effect of disorder on IR spectra**

As mentioned in the main text, the effect of local disorder or of thermal fluctuations may affect the IR spectra. This would explain the differences between the simulated and the experimental curves in Figure 2. In order to elucidate this point, we assume a simple model, namely a single, VDF infinite polymer, with one monomer per cell (Figure S6a). Although not catching the overall complexity of real nanostructures, such model allows the effect of the intrachain disorder to be highlighted in a controlled way. For instance, considering the effect of the rotation of the $CF_2$ units around the polymer axis, we calculate the IR spectrum for different angles of rotation (from 0° to 45°, with 0° corresponding to the configuration of minimum energy, assumed as the reference). In Figure S6a, the rotated $CF_2$ units are superimposed on the same structure, and distinguished with different colors (cyan=0°, red =15°, green=30°, blue=45°). For each system we calculate the phonon frequency at $\Gamma$ and the corresponding IR spectrum, along the line described in the main text. The results are shown in Figure S6b. The structural modification imparts a shits and a split of a few vibrational bands with respect to the ideal case (cyan line). As in general many internal angles may coexist we consider the average over these model configurations (black line). The final spectrum is broader and more structured than the ideal one (see e.g peak at ~1200 $cm^{-1}$), closely resembling the effects observed in experimental curves. However, we point out that the reference spectrum (cyan) already includes all the features of the system. Analogously, the simulated IR spectra in Figure 2 include all the spectroscopic features of the experimental systems.

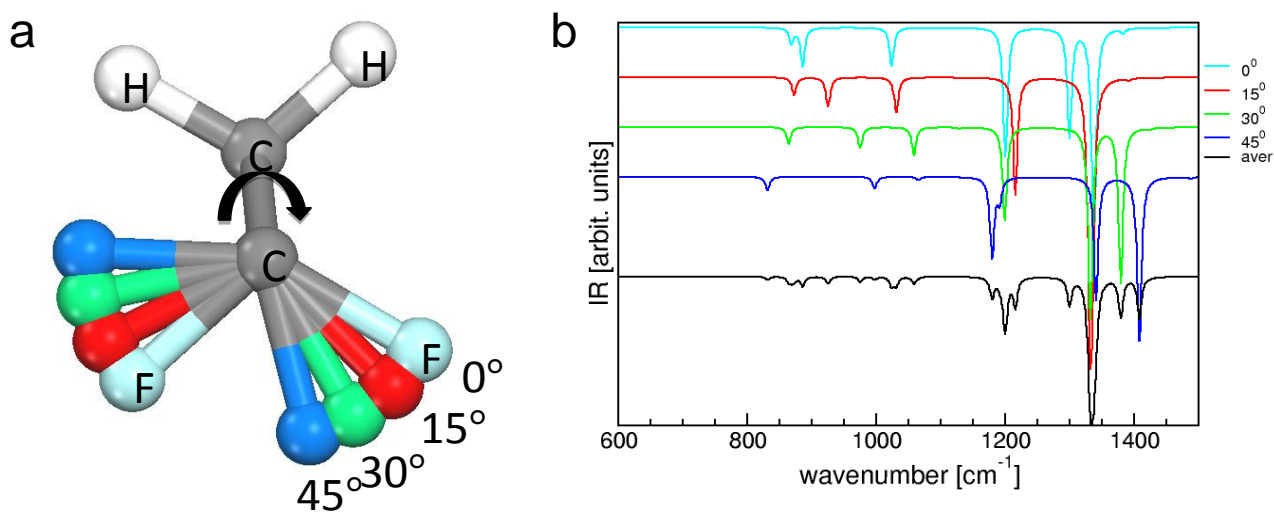

**Figure S6.** (a) Top view of a VDF monomer. The $CH_2$ unit is kept fixed in the minimum energy configuration, the $CF_2$ one is instead rotated around the polymer axis by 15°, 30°, 45° with respect to the minimum energy configuration (0°). (b) Simulated IR spectra for configurations corresponding to geometries of panel (a). Black line: average over the four considered configurations.

**References**

[S1] R. Resta, D. Vanderbilt, in *Physics of ferroelectrics: a modern perspective*. (Eds. C. H. Ahn, K. M. Rabe, J. M. Triscone, Springer-Verlag 2007, Ch. 2.

[S2] D. Vanderbilt, *J. Phys. Chem. Sol.* **2000**, *61*, 147.